\documentclass[aps,12pt,superscriptaddress,nofootinbib,floatfix,preprintnumbers]{revtex4}
\usepackage{amsmath}
\usepackage{graphicx}
\usepackage{amssymb}
\usepackage{amsfonts}
\usepackage{latexsym,epsfig}

\def\beq{\begin{eqnarray}}
\def\eeq{\end{eqnarray}}

\def\al{\alpha}
\def\be{\beta}

\def\ga{\gamma}

\def\vp{\varepsilon}

\def\na{\nabla}
\def\pa{\partial}

\begin{document}
\preprint{MSUHEP-061231}
\title{Torsion Phenomenology at the LHC}
\author{A.S. Belyaev}
\affiliation{Department of Physics and Astronomy,
          Michigan State University,
      East Lansing, MI 48824, USA}
\email{ belyaev@pa.msu.edu}
\author{I.L. Shapiro}
\affiliation{Departamento de F\'{\i}sica -- ICE,  
Universidade Federal de Juiz de Fora,  Juiz de Fora, 36036-330, MG,  Brazil 
\footnote{{Also at Tomsk State Pedagogical University, Russia}}}
\email{shapiro@fisica.ufjf.br}
\author{M.A.B. do Vale}
\affiliation{Departamento de Ci\^encias Naturais,  
Universidade Federal de S\~ao Jo\~ao del Rei, \
S\~ao Jo\~ao del Rei, 36301-160, MG, Brazil }
\email{aline@ufsj.edu.br}

\begin{abstract}
$\,$
\\
\\
We explore the potential of the CERN  Large Hadron Collider (LHC)  to test  the
dynamical  torsion parameters. The form of the torsion  action can be 
established from the requirements of consistency of  effective quantum field
theory. The most phenomenologically relevant  part of the torsion tensor is dual
to a massive axial vector field.  This axial vector has geometric nature, that
means it does not  belong to any representation of the gauge group of the SM
extension  or GUT theory. At the same time, torsion should interact with all 
fermions, that opens the way for the phenomenological applications.

We demonstrate that LHC collider can establish unique constraints  on the
interactions between fermions and torsion field considerably exceeding present
experimental lower bounds on  the torsion couplings and its mass. 
It is also shown how possible non-universal nature of torsion couplings due 
to the renormalization group running between the Planck and TeV energy 
scales can be tested via the combined analysis of Drell-Yan and $t\bar{t}$ 
production processes.
\end{abstract}
\newpage
\maketitle
\tableofcontents


\section{Introduction}

The Standard Model of particle physics (SM) is a very successful
theory, however there is an expectation that one day we observe 
the signs of a new physics beyond the scope of the SM. Which kind 
of a new physics can be expected? Answering this question is an 
important issue, because the same experimental data can be 
always interpreted in many different ways and only consistent 
theoretical consideration can provide a reasonable perspective
for a phenomenological interpretation of experimental data. 
From the theoretical viewpoint, one of the main shortcomings
of the Standard Model and its natural (non-minimal) extensions 
is their inability to incorporate quantum gravity. The natural 
conclusion is that, if the consistent quantum theory of all 
four fundamental interactions is really possible, it should 
include certain extension of the known physics, that is SM
plus General Relativity(GR). Modifications should be expected,
at the first place, in the gravitational sector of the theory. 

The most realistic candidate to be the universal theory of 
everything is the (super)string theory. The merit of this 
theory (along with very important advances in formal purely 
theoretical developments) is the possibility to incorporate 
gravity into the scheme of unification. All fundamental 
interactions emerge as low-energy manifestations of the 
fundamental quantum object - (super)string. The success of 
the superstring program produced a variety of mathematically 
perfect versions of this theory \cite{GSW,Polchinsky,Ketov}. 
Unfortunately there is no real perspective for any existing 
or projected experiment which could  give an answer which 
one of these mathematically perfect theories is actually 
correct. The situation becomes even more complicated if we 
remember that the consistent theory 
of superstring can be formulated only in higher dimensions
and that extra dimensions may be compactified in many 
different ways. Moreover at any finite order of the string
perturbation theory there is an additional ambiguity of 
defining the parametrization of external background fields 
\cite{zwei,Tseyt,Dere} (see also detailed discussion of this 
issue in \cite{torsi}). 

In this situation it is completely justified to apply a 
phenomenological approach and simply consider most natural 
extensions of GR assuming they might come from the unknown 
fundamental theory. The remarkable examples of such an 
approach are extra dimensions leading to TeV scale gravity 
\cite{extra dim} and Lorentz and CPT violations \cite{CPTL}. 
In this paper we shall concentrate on the phenomenological 
aspects of torsion gravity theory, which is traditionally 
considered as one of the most natural extensions of GR 
\cite{hehl-76,torsi}. Let us notice that global torsion 
field is in fact one of the main candidates for the 
Lorentz and CPT violation parameters.
Such field can be a component of vacuum resulting from 
some new symmetry breaking phenomenon \cite{torsi,Kost}.
Another possibility is to treat torsion as fundamental 
propagating field, which has well defined action and is 
characterized by such parameters as torsion mass and the 
values of the coupling between torsion and fermions (quarks 
and leptons). In the framework of effective quantum theory 
one faces rigid restrictions on the torsion parameters, 
originating from both phenomenological and theoretical analysis 
\cite{betor,guhesh,lebedev,Das:2002qv,mahanta}. In the present 
paper we study the LHC potential to improve limits on the 
torsion parameter space which we have derived earlier from LEP 
and Tevatron data~\cite{betor}.

The torsion field $T^\alpha_{\;\;\beta\gamma}$ is defined as 
follows\footnote{See \cite{hehl-76} and \cite{torsi} for 
different introductions to torsion}: 
$$
{\Gamma}^\alpha_{\;\beta\gamma} -
{\Gamma}^\alpha_{\;\gamma\beta} =
T^\alpha_{\;\;\beta\gamma}\,.
$$
It proves useful to divide torsion into irreducible components
\beq
T_{\alpha\beta\mu} =
\frac{1}{3} \left( T_{\beta}g_{\alpha\mu} -
T_{\mu}g_{\alpha\beta} \right)
- \frac{1}{6} \varepsilon_{\alpha\beta\mu\nu}
\,S^{\nu} + q_{\alpha\beta\mu}\,,
\label{t1}
\eeq
where the axial vector \ $S^\mu$ \ is dual to the completely 
antisymmetric torsion tensor, $T_\al$ is a vector trace of torsion 
and \ $q_{\alpha\beta\mu}$ \ is a tensor which satisfies the 
constraints  \ $q^\al_{\,\cdot\,\be\alpha}=0$ \ and
  \ $q_{\alpha\beta\mu}\vp^{\alpha\beta\mu\nu}=0$.

The general nonminimal action of a Dirac fermion coupled to torsion
has the form
\beq
S_f \,=\, \int\sqrt{g}\,\left\{\,i\bar{\psi}\gamma^\mu 
\big( \na_\mu - i \eta_1\gamma^5S_\mu + i\eta_2T_\mu\big)\psi 
- m\bar{\psi}\psi \right\}\,,
\label{t2}
\eeq
where $\eta_1,\,\eta_2$ are nonminimal parameters and 
$\na_\mu$ is Riemannian covariant derivative (without 
torsion). Let us remember that the minimal interaction 
corresponds to the action 
\beq
S_{min,f} \,=\, \frac{\imath}{2}\int d^4x\sqrt{-g}\,\left\{\,
\bar{\psi}\ga^\mu {\tilde \na}_\mu\psi 
- {\tilde \na}_\mu\bar{\psi}\,\ga^\mu\psi
\,+\, 2\imath\, m\,\bar{\psi}\psi\,\right\}\,,
\label{minimal action}
\eeq
where \ ${\tilde \na}_\mu$ \ is the covariant derivative 
{\it with} torsion. Direct calculation shows that this expression
corresponds to the values $\eta_1 = -1/8,\;\;\eta_2 = 0$ \ of 
parameters in (\ref{t2}). It has been shown long ago
\cite{bush85} that the quantum theory is not renormalizable
for a fixed non-zero value of $\eta_1=\eta$, while the absence of 
$\eta_2$ does not imply any special difficulties. Therefore, 
in what follows we consider $\eta_1$ as an arbitrary parameter 
and take $\eta_2=0$ for simplicity. This is equivalent to 
consider the completely antisymmetric torsion tensor
$T_{\al\be\ga}= - \frac{1}{6} \varepsilon_{\alpha\beta\ga\mu}
\,S^{\mu}$. So, we shall choose the following action for 
each of the fermions $\psi_{(i)}$:
\beq
{\cal S}_{non-min}^{TS-matter} 
\,=\, i\,\int d^4x\sqrt{g}\,\,{\bar \psi}_{(i)}\,
\Big(\, \ga^\al\,\na_\al  + \imath\,\eta_{i}\ga^5 \ga^\mu S_\mu
- \imath\, m_{i} \,\Big)\,\psi_{(i)}\,,
\label{dirac2-nm}
\eeq
where $\eta_{(i)}$ is the non-minimal interaction parameter
for the corresponding spinor. Let us remember that torsion 
is a tensor which does not depend on metric. One can always 
assume, for simplicity, that the metric is flat 
\ $g_{\al\be}=\eta_{\al\be}$ \ and therefore the torsionless 
part of the covariant derivative is nothing else but a partial 
derivative \ $\na_\al=\pa_\al$.

The action for the propagating torsion can be 
established using the consistency criteria for the effective
low-energy quantum theory of torsion coupled to the gauge 
model with fermions and scalars, such as the Standard Model.  
Requesting unitarity and renormalizability of the theory in 
the low-energy sector one arrives at the unique possible form 
of the torsion action 
\cite{betor}
\beq
{\cal S}_{tor}^{TS-kin}\,=\,\int d^4 x
\,\left\{\,-\frac14\,S_{\mu\nu}S^{\mu\nu}
+ \frac12\,M_{TS}^2\, S_\mu S^\mu\,\right\}\,,
\label{action}
\eeq
where $M_{TS}$ is the mass of the torsion and 
$\,S_{\mu\nu}=\pa_\mu S_\nu-\pa_\nu S_\mu$. 
The consistency of the theory based on the action (\ref{action})
holds when the quantum effects of the fermion loops are taken 
into account. After that point one can perform both theoretical 
and phenomenological analysis\footnote{Similar phenomenological 
analysis based on the longitudinal torsion has been performed 
in \cite{Carroll}. However the longitudinal torsion is 
inconsistent from the formal viewpoint.}. 

The first constraints on the dynamical torsion parameters have 
been established in \cite{betor} based on LEP1.5 and Tevatron 
data for the theory with the action (\ref{action}).
These constraints have been updated in later paper~\cite{mahanta}
using improved LEP2 and Tevatron statistics.
In \cite{lebedev} constraints for the similar theory with extra 
dimensions have been presented. Furthermore, other phenomenological 
and theoretical aspects of torsion, in the framework of high 
energy theory, have been explored in \cite{alim} and \cite{freidel}.

The theoretical investigation which is the most important for us 
has been performed in \cite{betor,guhesh}, where the consistency of 
the effective theory with quantized torsion was studied. Following 
the standard approach, we postulate that all fields have to 
be quantized in the framework of effective approach, and 
found that, for the torsion-fermion system this can be 
consistently done only if the constraint \ $M_{TS}/\eta_i \gg m_i$ 
( or, equivalently, $M_{TS} \gg \eta_i \times m_i$) is 
satisfied for all fermions with the masses $m_i$ and with the 
nonminimal parameter \ $\eta_i$.
Despite 
this restriction does not rule out the existence 
of torsion, it puts a strong constraint either on its 
propagation or on its interactions with fermions. Together
with the phenomenological bounds for the two parameters 
$M_{TS}$ and $\eta$ \cite{betor} we arrive at the very strong 
constraints on the  torsion parameters. The purpose of the 
present article is to study the LHC potential to put further 
constraints on parameters $M_{TS}$ and $\eta_i$. As we show,
the limits on the ($M_{TS},\eta$) parameter space which can 
be obtained at the LHC, are much more restrictive than the 
ones obtained previously using LEP and TEVATRON  data \cite{betor}.

The paper is organized as follows. In the next section we 
shall briefly discuss the theoretical background of effective
quantum theory of torsion. In this part we assume the reader can 
consult the previous publications \cite{book,betor,guhesh,torsi} 
for more detailed and pedagogical introduction. Furthermore, we 
shall calculate the approximate relations between the 
parameters $\eta_i$ for different kinds of fermions $\,i$, using 
renormalization group and a very natural assumption of equal 
$\eta_i$ at the Planck scale, where all fields should be generated 
and start to interact with the geometry described, in particular, 
by torsion. In section 3 we shall discuss the phenomenology
of torsion for $pp\to \ell\ell$ and $pp\to t\bar{t}$
processes at the LHC and compare the corresponding upper bounds 
for the torsion parameters \ $M_{TS}$ \ and \ $\eta$ \ with the 
ones for obtained previously from LEP and Tevatron. 
Finally, in section 4 we draw our conclusions. 

\section{The Effective Approach to Torsion\label{sec2}}

The considerations which lead to the antisymmetric torsion 
action (\ref{action}) look as follows. As far as torsion is 
considered as a dynamical field, one has to incorporate it into 
the SM along with other vector fields. Due to the assumed 
geometric (gravitational) nature of torsion, it is coupled to 
all fermionic fields. The action of these fields looks as 
(\ref{dirac2-nm}) with additional SM-type interaction to 
the gauge and scalars fields.

The fermionic sector of the gauge theory consists of the 
actions (\ref{t2}) supplemented by the gauge and Yukawa 
interactions. The symmetries of the theory include usual gauge 
transformations and an extra transformation related to torsion
\beq
 \psi \to \psi^\prime = \psi\,e^{\ga_5\be(x)}
,\,\,\,\,\,\,\,\,\,\,\,\,\,\,
{\bar {\psi}} \to {\bar {\psi}}^\prime 
= {\bar {\psi}}\,e^{\ga_5\be(x)}
,\,\,\,\,\,\,\,\,\,\,\,\,\,\,
S_\mu \to S_\mu^\prime = S_\mu - {\eta}^{-1}\, \pa_\mu\be(x)\,.
\label{gamma 5}
\eeq
Massive fermion term is not invariant under the last transformation,
so it corresponds to the softly broken symmetry. This symmetry has 
a significant impact on the renormalization structure of the theory. 
Since it is softly broken, it does not 
forbid massive counterterms in the torsion sector. Therefore,
$S_\mu$ has to be a massive field, for otherwise one can not 
control quantum corrections to the corresponding massive term. 
Furthermore, the renormalization in the massless sector does 
not depend on the massive parameters. If we start from the most 
general action with second derivatives for \ $S_\mu$,
\beq
S_{tor} = \int d^4 x\,\left\{\, 
-a\,S_{\mu\nu}S^{\mu\nu} + b\,(\pa_\mu S^\mu)^2
+ \frac12\,M_{ts}^2\,S_\mu S^\mu\,\right\}\,,
\label{geral1}
\eeq
an extra symmetry (\ref{gamma 5}) tells us that the 
\ $S_{\mu\nu}S^{\mu\nu}$-type counterterms really takes 
place. Direct calculations also confirm this fact and hence 
the $\,-a\,S_{\mu\nu}S^{\mu\nu}$-term is necessary in the 
consistent theory for the same reason as the massive term. 
If we do not introduce the transverse kinetic term, it will 
emerge at quantum level anyway. Furthermore, the co-existence 
of the $-a\,S_{\mu\nu}S^{\mu\nu}$-term and of the
\ $b\,(\pa_\mu S^\mu)^2$-term violate unitarity of 
the theory. In this way we arrive to the unique form of 
the torsion action (\ref{action}).  

The high energy phenomenology of the torsion theory 
(\ref{action}) is based on the fact the axial vector 
\ $S_\mu$ \ has no direct analogs in the Standard Model. 
At the first place, this is a massive axial vector and, 
also, it is not related to a gauge group of the theory and 
interacts with the matter fields only via the \ $\eta$-coupling 
in the Eq.~(\ref{dirac2-nm}). At the same time, this interaction 
is not necessarily universal
since the values of parameters 
\ $\eta_i$ \ may be different for distinct fermions. The 
renormalization group equations for different \ $\eta_i$ \ 
are dependent on the corresponding Yukawa couplings
\cite{bush85,betor}. As far as these couplings are different, 
the renormalization group equations for different \ $\eta_i$ \ 
are also different and finally the values of these parameters 
at the low-energy scale should be also different. 

The renormalization group equations for \ $\eta_i$ \ have
been obtained and discussed in \cite{betor} in a rather 
general form. In order to evaluate the values of  \ $\eta_i$ \ 
at the given scale we shall assume that (as it was already 
mentioned before) the values of all \ $\eta_i$ \ are equal at
the Planck scale. In this case the set of the values of the 
nonminimal parameters \ $\eta_i(\mu_{LHC})$ \ at the energy 
scale available in LHC experiments should result from the 
renormalization group flow for these parameters. Let us 
start from the simple version of the renormalization group 
equations \footnote{We correct here the misprint of 
\cite{betor}.}
\beq
(4\pi)^2\,\mu\,\frac{d\eta_i(\mu)}{d\mu}
\,=\,\frac13\,\eta_i^3(\mu) + C\,h_i(\mu)\,\eta_i(\mu)\,.
\label{RG for eta}
\eeq
The solution of this equation is simple provided we know
the function $\,h_i(\mu)$, but in fact this function is 
available only for relatively low energies where the SM 
can be experimentally verified. At higher energy scale 
one needs to know a corresponding gauge model (being it 
some GUT or supersymmetric extension of the SM or 
something else, e.g. just a proper SM valid until the 
Planck scale). Of course, this information is not available.
However, as we shall see in a moment, one can obtain the
relevant data even without it. 

Let us start from the minimal assumptions that the 
effective Yukawa coupling are constants which possess
the same values as all scales. Furthermore we
assume a very small $\eta_i$, such that only the second
term in the Eq.~(\ref{RG for eta}) is relevant. Then
the solution of the above equation becomes very simple 
\beq
\eta_i(\mu)\,=\,\eta_i(\mu_0)\,
\Big(\frac{\mu}{\mu_0}\Big)^{Ch_i/(4\pi)^2}\,.
\label{scale eta}
\eeq
Our working hypothesis is that the nonminimal parameters 
are $\,\eta_i(\mu_0)\equiv\eta_{Planck}\,$ equal at the 
fundamental Planck scale $\mu_0=M_P=10^{19}\,$~GeV. 
Then the values of $\,\eta_i(\mu)\,$
depend only on the expression \ $Ch_i/(4\pi)^2$. Let us take 
an extreme case of the top quark, when $\,h_i\sim 1$ and 
assume, as it was already discussed before, that this value 
does not depend on scale. The coefficient $\,C\,$ has been 
derived for various models based on different gauge groups 
\cite{bush85,book}.
A very optimistic estimate is $\,C=5\,$, that requires 
a sufficiently large gauge group of GUT. Then we obtain,  
for $\,\mu_0=M_P=10^{19}\,GeV\,$ and $\,\mu=10^{3}\,GeV$,
the estimate  
\beq
\Big(\frac{\mu}{\mu_0}\Big)^{Ch_i/(4\pi)^2}\approx 
10^{-1/2}\,.
\label{eta mu}
\eeq
Obviously, due to our estimate \ $h_i\sim 1$, this 
result has relation only to the top quark case, while
for other fermions the ratio will be very close to unity. 
Now, let us notice that in all our simplifications we 
have always made a choice in such a way that the ratio in 
(\ref{eta mu}) was becoming bigger. For instance, already 
for SM the value of $C$ is much smaller than $5$, the running 
of the Yukawa constants, being in accordance with the asymptotic 
freedom, reduces the effect and finally taking into account 
the nonlinear term in Eq.(\ref{RG for eta}) is also decreasing 
the ratio in (\ref{eta mu}). All in all, we conclude that, 
except the case of the top quark, the values of all $\,\eta_i$ 
at the TeV scale must be equal or at least very close to 
each other, in fact they are simply equal to the universal 
value $\eta_{Planck}$. The top quark parameter $\,\eta_t\,$
may be, in principle, different from the others because
of the potentially stronger running between the Planck 
and Fermi (or TeV) scales. The lesson we learned here is 
that the list of unknown parameters of the theory includes 
the mass of the torsion $\,M_{ts}\,$ plus the universal  
nonminimal parameters $\eta_{i\neq t}\equiv\eta$ and also 
$\eta_t$ which may be a bit smaller than $\eta$. In the previous 
papers on the subject \cite{betor} we did not take the possible 
difference between $\eta$ and $\eta_t$ into account.

In the present case we are going to take into account the 
processes involving the top quark and also the ones involving 
other fermionic particles. Our purpose is to improve the upper
bound for the torsion parameters and therefore it is reasonable 
to account for the possible difference between $\,\eta_t\,$ and 
$\,\eta_{i\neq t}\equiv \eta$.

\section{Phenomenological Consequences at LHC}

The CERN Large Hadron Collider (LHC) in about one
year  will start proton-proton collisions at 14 TeV in
the center of mass energy. Thanks to its unprecedented
energy and luminosity, the LHC will be able to probe
the new physics at the several TeV scale.

The straightforward consequence of the heavy torsion
interacting with fermion fields is the effective
four-fermion contact interaction of leptons and quarks 
(see details in Ref.~\cite{betor}). Since torsion
interacts universally with light quarks and leptons, the
most sensitive process  to probe  torsion  interactions
with light fermions is the Drell-Yan (DY)  process of
dilepton production.
It has been shown in ATLAS and CMS collaborations studies
that compositeness scale
at the LHC 
can be probed down to 20-30 TeV scale via 
DY process~\cite{atlas-tdr,cms-tdr}.
On the other hand, the $pp\to t\bar{t}$
process is the natural choice
to probe torsion coupling with top-quark.

To perform our analysis we have implemented
interactions of torsion with fermions described
by Eq.~\ref{dirac2-nm} into CalcHEP package~\cite{calchep}.
The study has been done at the parton-level
using CalcHEP  2.45 together with the realistic experimental 
efficiencies. The CTEQ6M parton distribution function
(PDF) with the QCD renormalization and factorization scales 
equal to the torsion mass $M_{TS}$ was used in CalcHEP.
In our analysis we assume a total integrated luminosity of 
100 fb$^{-1}$.

\subsection{Limits from di-lepton production}
\begin{figure}
\includegraphics[width=0.5\textwidth]{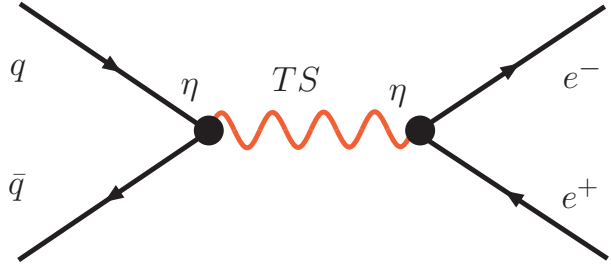}
\caption{Additional Feynman Diagram for process 
$\,p p \longrightarrow e^+ e^-\,$ at LHC introduced 
by the torsion (TS) interactions.}  
\label{fig:diag1}
\end{figure}

\begin{figure}
\includegraphics[width=0.7\textwidth]{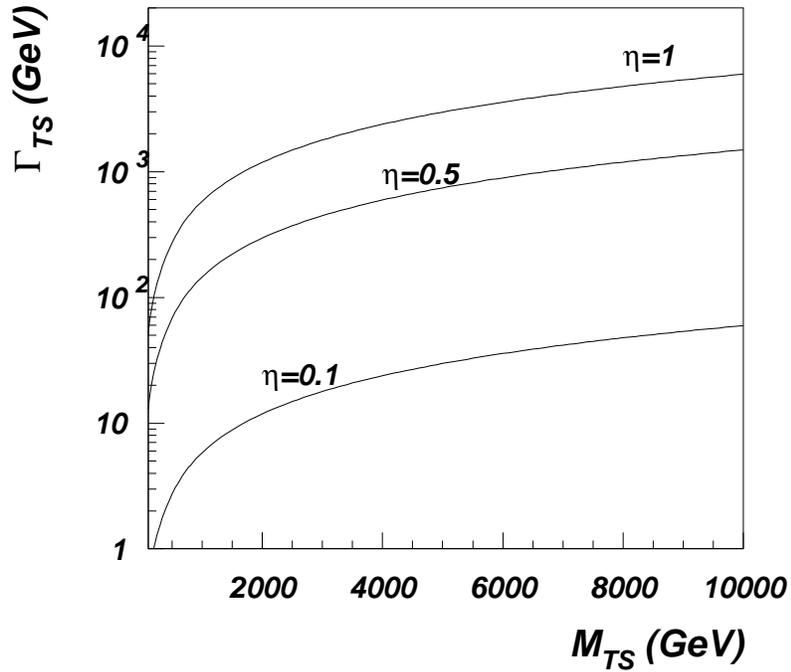}
\caption{
Torsion width versus its mass for various
values of $\eta$ coupling
under assumption $\eta=\eta_{top}$  \label{fig:tsw}}
\end{figure}

\begin{figure}
\includegraphics[width=0.7\textwidth]{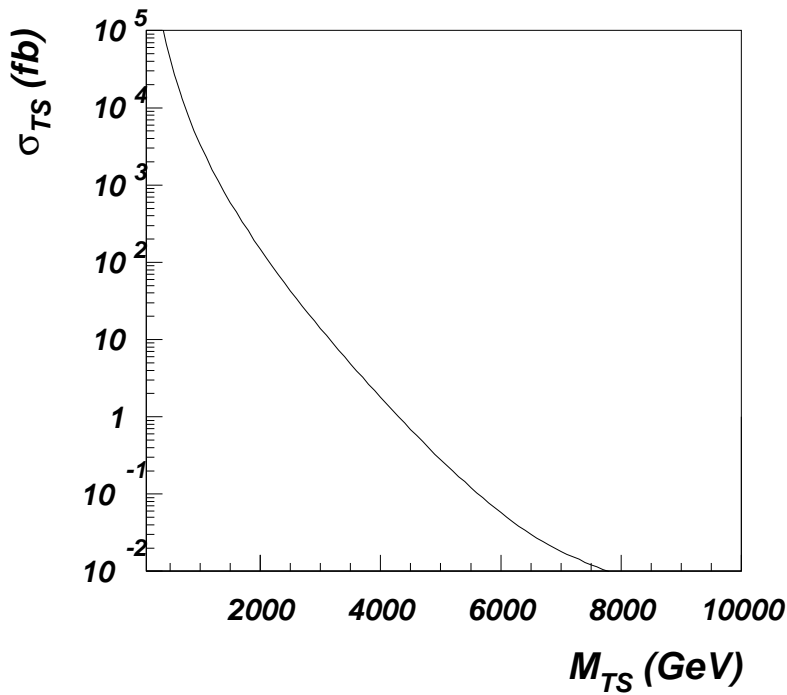}
\caption{
 Torsion production cross section at the LHC for $\eta=0.1$.
\label{fig:csts}}
\end{figure}

In Fig.~\ref{fig:diag1} we present Feynman diagram
for torsion contribution to di-lepton production.
One should note
that heavy torsion can not be exactly approximated
by four-fermion contact interactions in general,
since it has a significant width, as one can see from Fig.~\ref{fig:tsw}.
For  small values of $\eta$ coupling
resulting to $M_{TS}\gg\Gamma_{TS}$
four-fermion contact interaction
would be a good approximation
for  $\hat{S}< M_{TS}$.
In our studies 
we use exact interaction described by Eq.~(\ref{dirac2-nm})
and do not  use approximation
of the contact interaction of torsion.
In this way we take into account both cases --
when torsion is 
being produced either on-shell or off-shell.
The cross section of torsion production rapidly 
drops down to about $10^{-1}$fb with the increasing of its mass up to about 
5 TeV for $\eta=0.1$ as one can see in Fig.~\ref{fig:csts}.
For this cross section one could expect about 10 events from the torsion
which would be observable at the LHC in case if background
reduced down enough with the appropriate choice of the kinematical cuts.
The cross section of the torsion produced on-shell
scales as $\eta^2$ and as $\eta^4$
in case of the  off-shell  torsion production.

To take into account  the detector acceptance we require the
rapidity of the electron and the positron to be  between -2.5 and
2.5. We have also found that the cut $p_T^e>M_{TS}/3$ optimizes
the signal significance together  with the mass window cut on 
the $e^+e^-$ invariant mass: $|M_{TS}-M_{e^+e^-}|< 2 \times \Gamma_{TS}$,
where $\Gamma_{TS}$ is the torsion width.

In our study we require signal-to-background ratio
(S/B) to be  bigger than 1/2
which happens to be always the case
after we apply kinematical cuts above.
This S/B ratio cut was
applied  to ensure the signal observation in the presence of
the possibly large (~50\%)  uncertainty in the parton density function
(PDF) in the high $x$ region of the heavy torsion production. 
The $P_T$ and mass window cut which we have
chosen actually provide good S/B ration above 1/2 and very
effective background suppression while leaving intact more
than 50\% of the signal events.

We have calculated the significance ($S/\sqrt{B}$) from the
signal+background (S+B) events excess over the background (B).
Under S+B we understand the rate of the signal+background
including  the effect of the signal and background
interference i.e. interference  between $\gamma/Z-$boson and
the torsion exchange. In the region of low statistics when the
number of signal events drops below 30, we have used Poisson 
statistics formula  for significance calculation.

\begin{figure}
\includegraphics[width=0.8\textwidth]{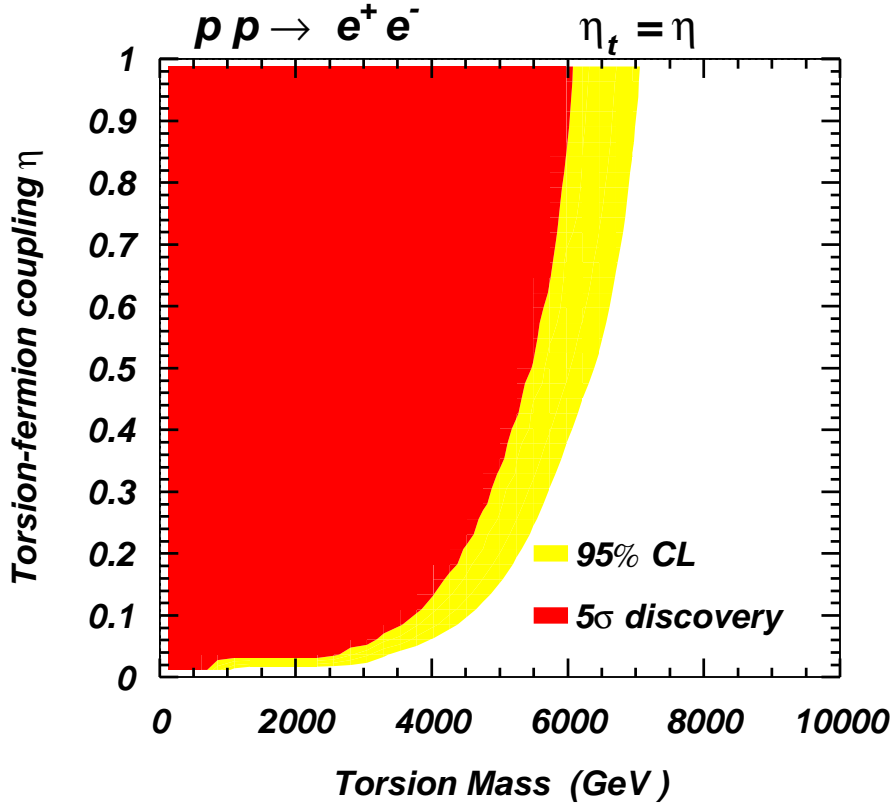}
\caption{The LHC discovery reach and 95\% CL 
exclusion bounds from $pp\to e^+ e^-$
process for the ($M_{TS},\eta$) parameter space.
\label{fig:pp_ee}}
\end{figure}

The LHC discovery reach and 95\% CL 
exclusion bounds from $pp\to e^+ e^-$
process for the ($M_{TS},\eta$) parameter space
are shown
in Fig.~\ref{fig:pp_ee}. 
One can see that for $\eta=0.1$, LHC can exclude 
$M_{TS}\lesssim 4.5$~TeV at 95\% CL while torsion with 
about 3.5~TeV mass can be discovered for this value of $\eta$. 
For $\eta=1$ LHC is sensitive to  about 7 TeV torsion at  
95\%CL while torsion with about 6 TeV mass can be discovered.

\begin{figure}
\begin{center}
\includegraphics[width=0.5\textwidth]{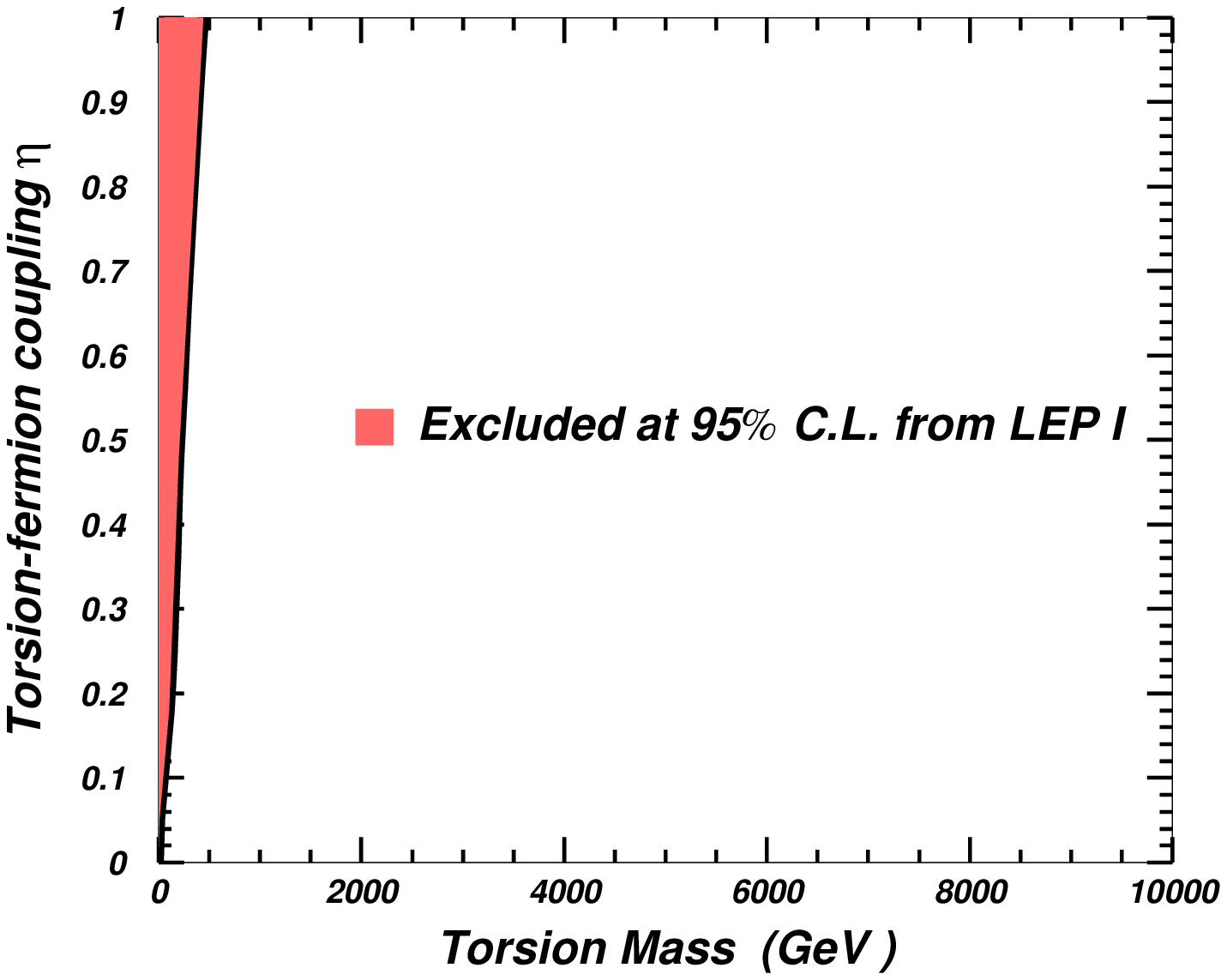}%
\includegraphics[width=0.5\textwidth]{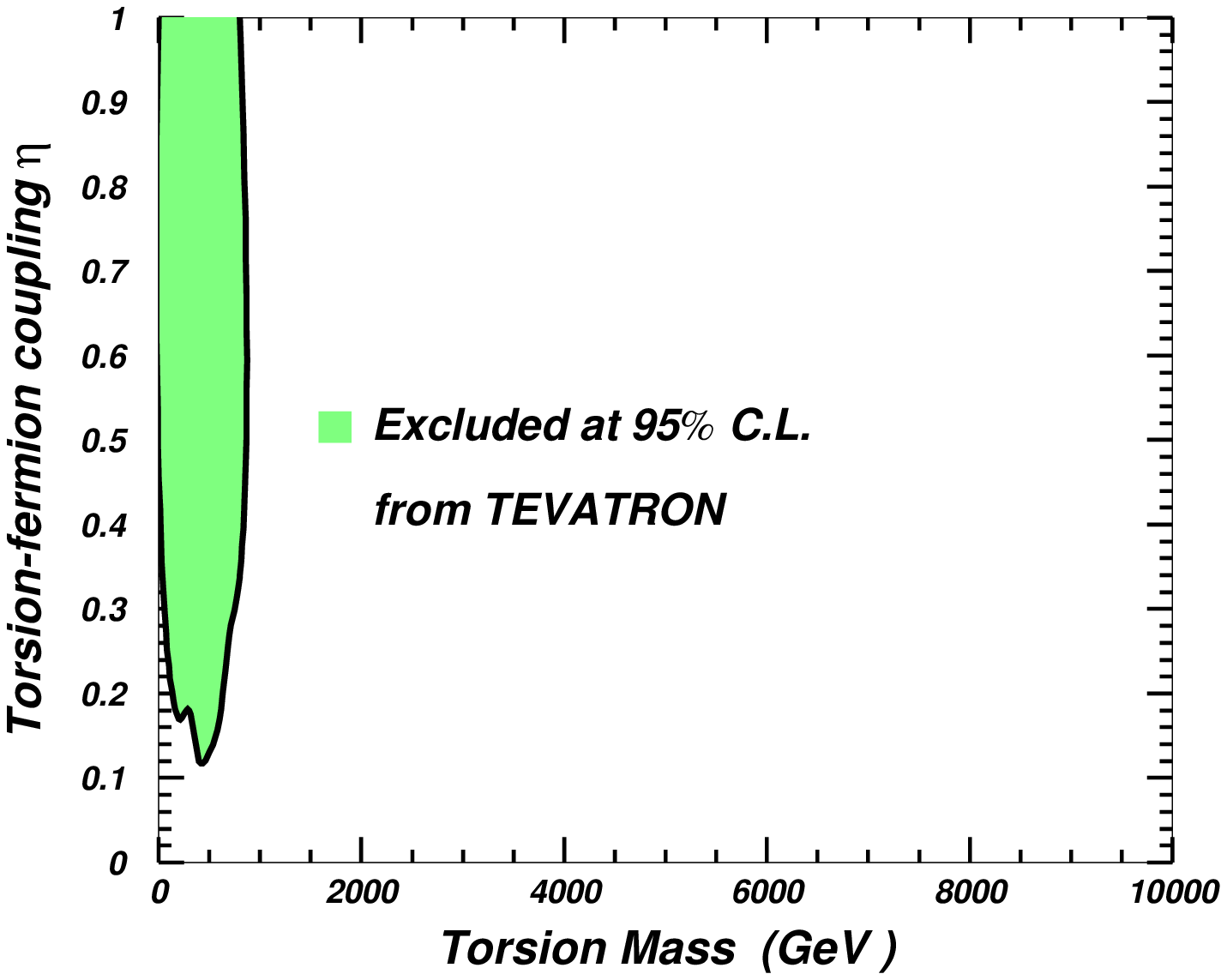}
\caption{Bounds for the ($M_{TS},\eta$) parameter space at  95\% C.L. 
            obtained from
	     from LEP data \cite{betor} (left) and
	     TEVATRON data \cite{betor} (right).\label{fig:betor_bounds}} 
\end{center}
\end{figure}

Comparing with previous constraints  from Tevatron and LEP on
the torsion parameters~\cite{betor} (Fig.~\ref{fig:betor_bounds}) 
(see also ~\cite{mahanta} for the improved LEP2 and Tevatron
bounds and Fig.5 in there)
we can see that indeed LHC is
much more sensitive to ($M_{TS},\eta$) parameter space. For a
given value of $\eta$, LHC could extend the torsion mass reach
by about one order of magnitude from di-lepton production.
Moreover, the more sophisticated analysis and combination of
electron and muon channels could improve further the LHC reach
for the torsion parameter space.

\subsection{Limits from $pp\to t\bar{t}$ process: the test of $\eta_t$ coupling}

As we pointed out in Section~\ref{sec2},
the $\eta_t$ coupling of torsion to top-quark
may not be necessarily universal
as compared to torsion couplings to the light fermions.

\begin{figure}
\includegraphics[width=0.5\textwidth]{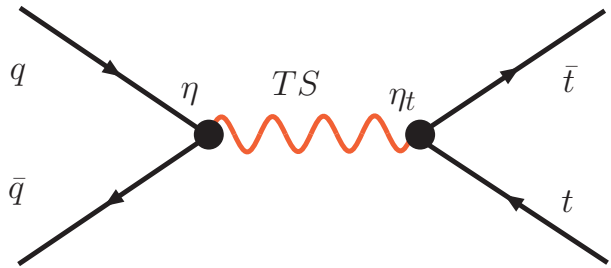}
\caption{Additional Feynman Diagram for process 
$p p \longrightarrow t \bar{t}$ at LHC 
introduced by the torsion interactions.} \label{fig:diag2}
\end{figure}

Since this coupling is the new additional parameter in low energy theory,
it should be tested on the same foot as the other parameters of the model.
Eventually, the  $pp\to t\bar{t}$
process is the naturally best choice
to probe torsion coupling to the top-quark.
In Fig.~\ref{fig:diag2} we present Feynman diagram
for
$p p \longrightarrow t \bar{t}$ 
process at LHC 
due to  the torsion exchange.
\begin{figure}
\includegraphics[width=0.5\textwidth]{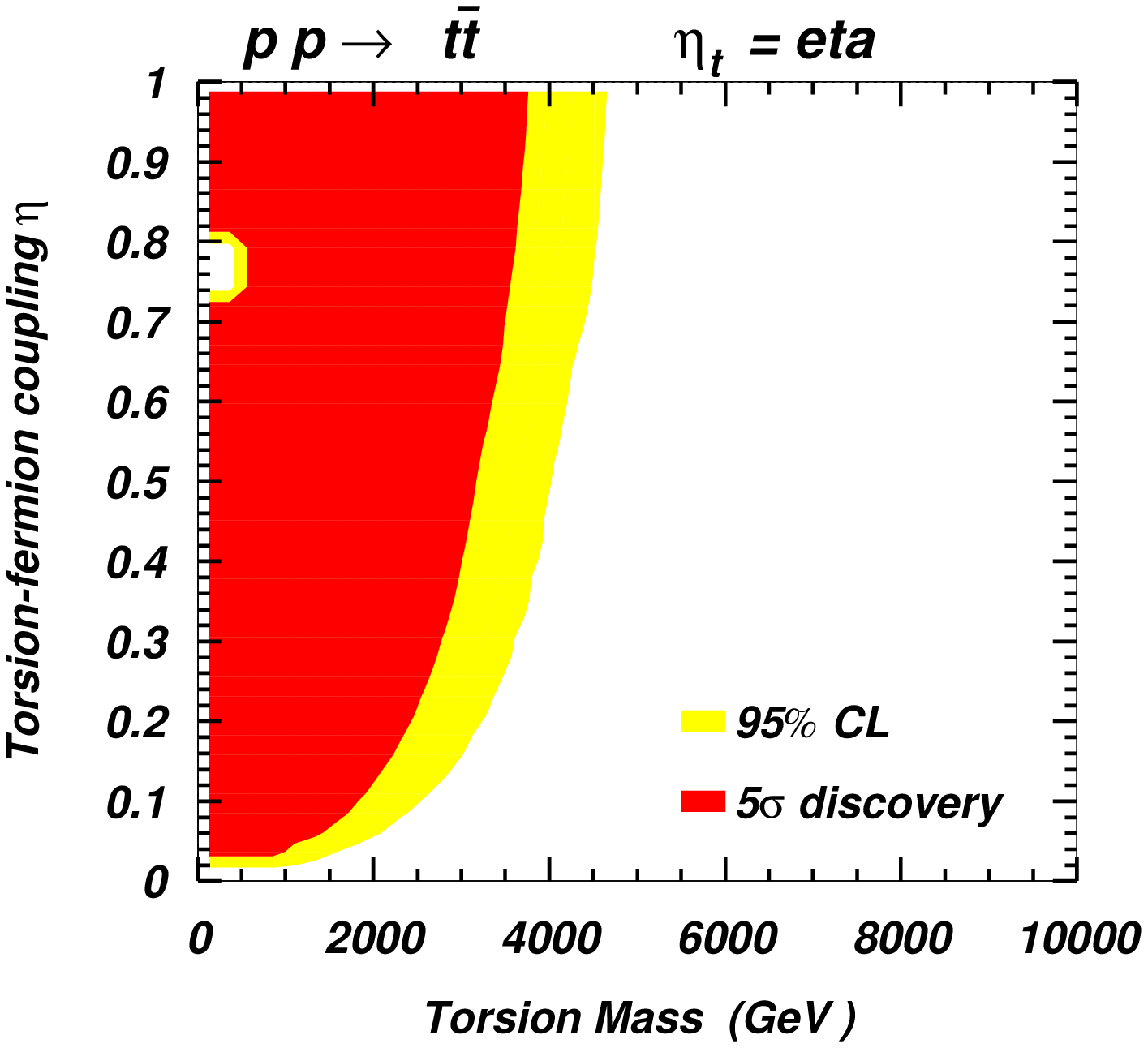}%
\includegraphics[width=0.5\textwidth]{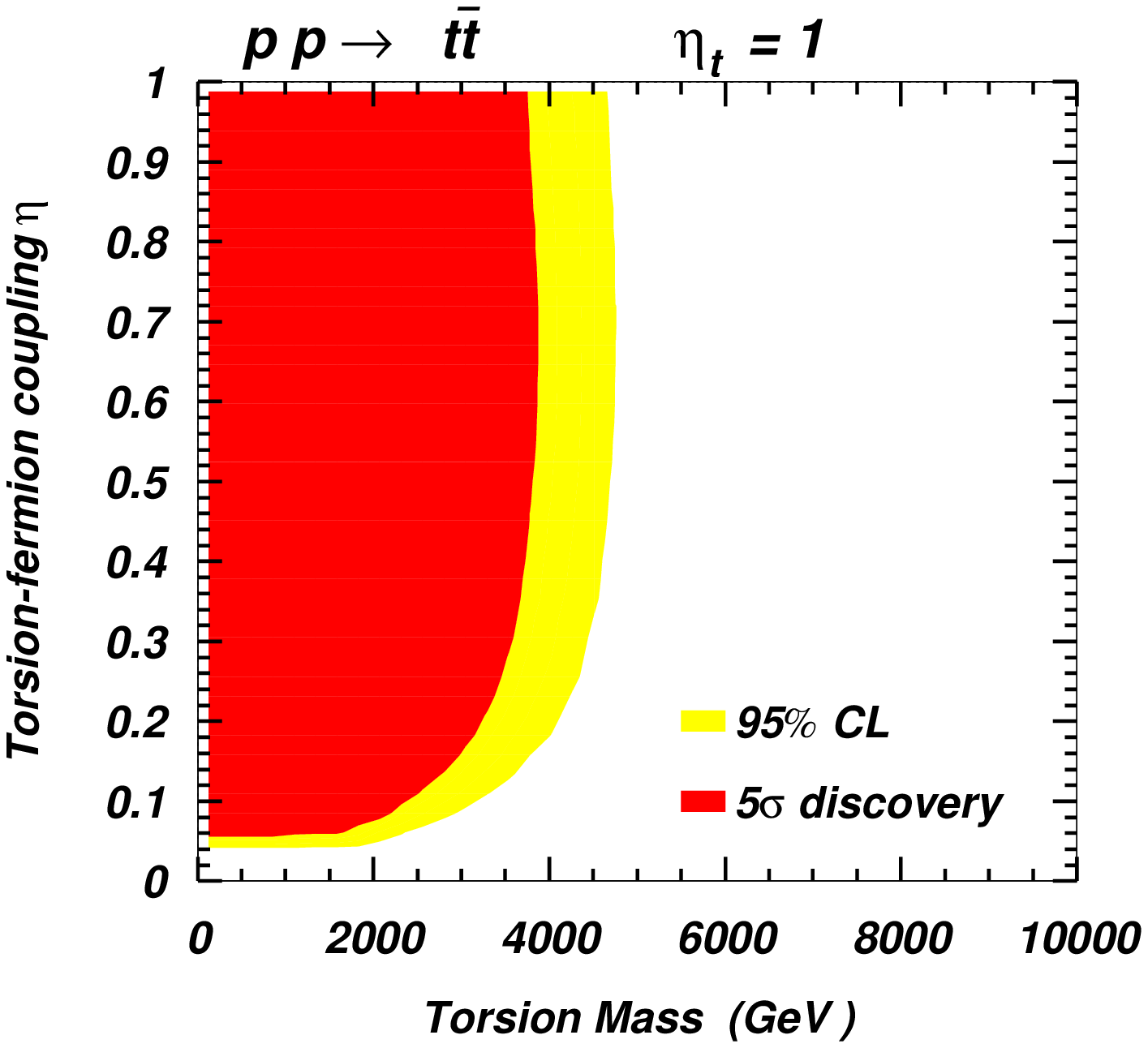}\\
\includegraphics[width=0.5\textwidth]{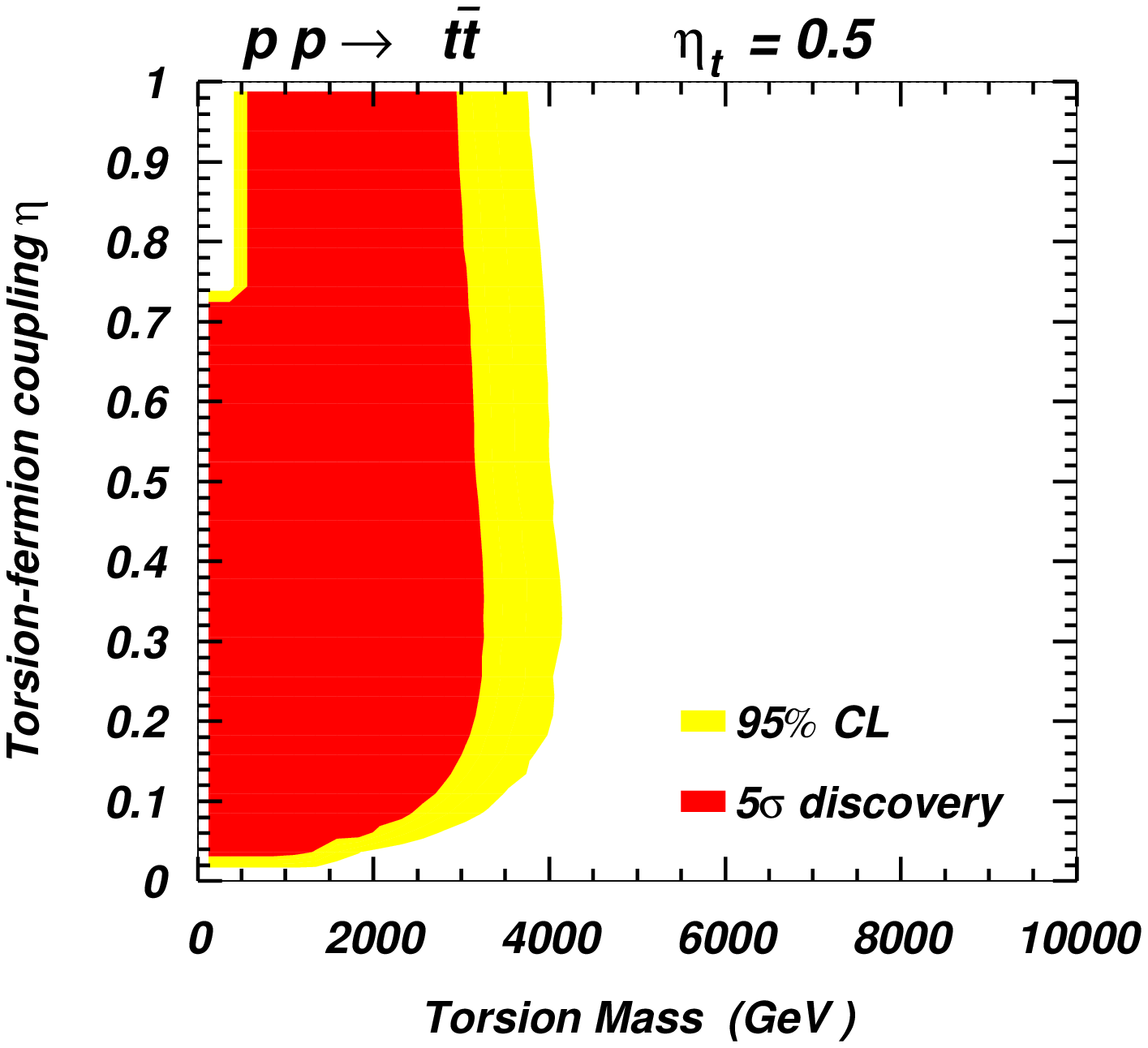}%
\includegraphics[width=0.5\textwidth]{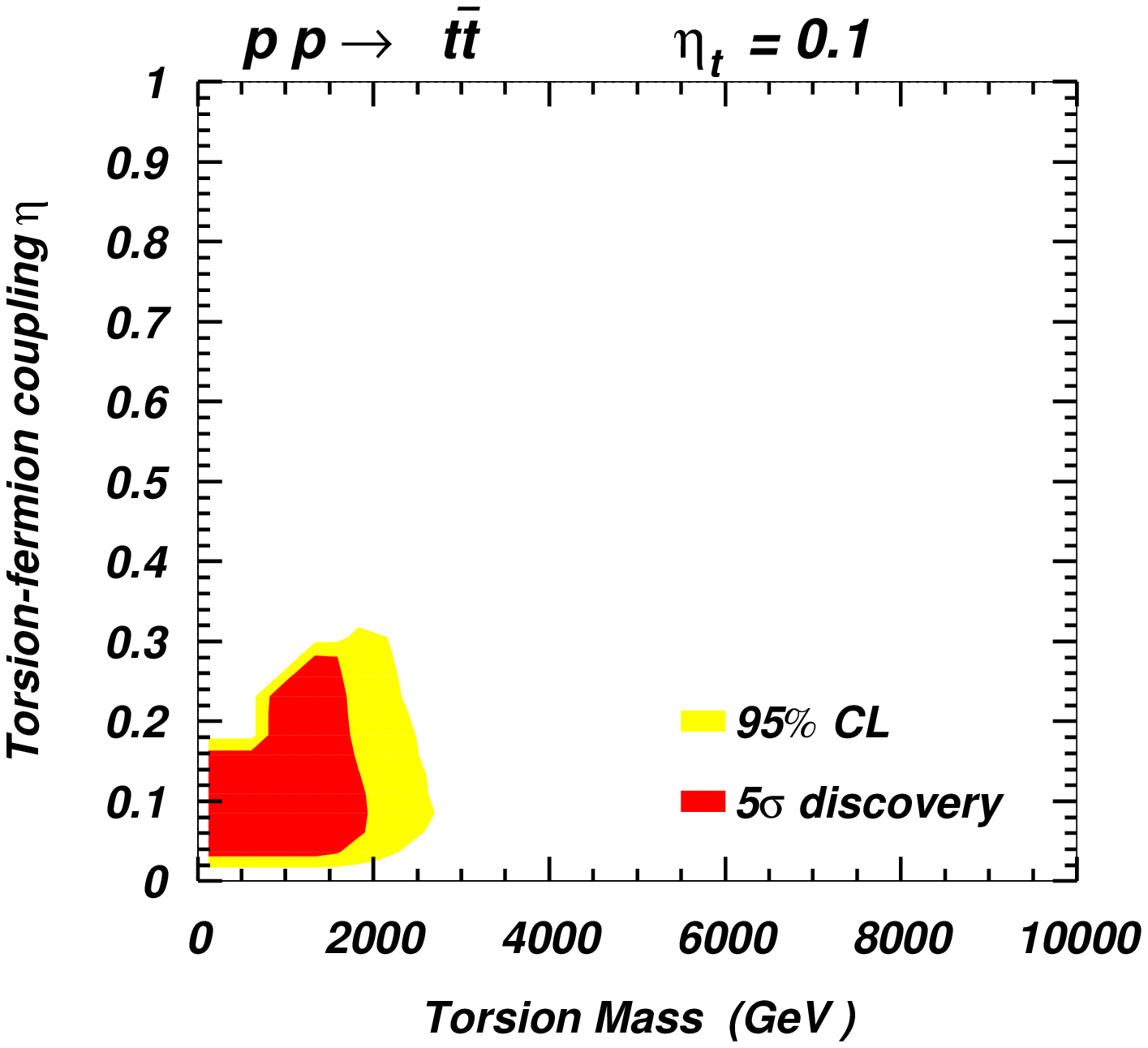}
\vskip -14.5cm
\hspace*{-7cm}a)\hspace*{8cm}b)
\vskip 6.5cm
\hspace*{-7cm}c)\hspace*{8cm}d)
\vskip 6cm
\caption{LHC sensitivity to ($M_{TS},\eta$) parameter space
for various cases of $\eta_t$ parameter:
$\eta_t=\eta$,
$\eta_t=1$,
$\eta_t=0.5$ and
$\eta_t=0.1$. \label{fig:pp_tt}}
\end{figure} 

Top-quark undergoes subsequent decay chains and should be
reconstructed. To keep the parton-level spirit of this 
study and correctly take into account qualitative
experimental effects, we have utilized the  1\%
efficiency of  the reconstruction of the $t \bar{t}$
pair. This efficiency has been obtained in the recent
studies~(see \cite{kkg} and reference [13] therein)
for the lepton-hadron decay
signature  of $t\bar{t}$ pair.

Similar to $pp\to e^+e^-$
process we use
$p_T^t>M_{TS}/3$ and  $|M_{TS}-M_{t\bar{t}}|< 2 \times \Gamma_{TS}$
kinematical cuts on the reconstructed top-quarks
to optimize the signal significance.
As a result of the analysis of $ p p \to t \bar{t}$
processes in Fig.~\ref{fig:pp_tt}
we present LHC sensitivity to ($M_{TS},\eta$) parameter space
for various choices of $\eta_t$ parameter:
$\eta_t=\eta$,
$\eta_t=1$,
$\eta_t=0.5$ and
$\eta_t=0.1$.

One can see that the sensitivity of the LHC to
torsion parameter space for  $p p \to t \bar{t}$
process is worse as compared to  $p p \to e^+e^-$.
It is not surprising, since   there are two main factors
which cause this reduction:\\
1) the top-quark pair reconstruction efficiency is only 1\%;\\
2) the relative $t\bar{t}$ QCD background 
   originating from the strong interactions is bigger as compared
  to the electroweak background for di-lepton production.\\
As a result, we have limited sensitivity of the LHC
to   the torsion parameters via   $p p \to t \bar{t}$ process. 
For example, for $\eta\simeq\eta_t=0.1$ case (Fig.~\ref{fig:pp_tt}a)
the $5\sigma$ sensitivity of  LHC to $M_{TS}$ for $pp\to t\bar{t}$
process
is reduced down to about 2 TeV 
(as compared to 3.5 TeV  $M_{TS}$
 for $5\sigma$ LHC for $pp\to e^+e^-$ process, 
 see Fig.~\ref{fig:pp_ee}).
Nevertheless,
the $pp\to t\bar{t}$ process provides a unique opportunity
to test the specific $\eta_t$ coupling.
Further comments on  Fig.~\ref{fig:pp_tt}
are in order.
 In case of $\eta_t<\eta$
 as one can see in Fig.~\ref{fig:pp_tt}c,d
 the  LHC sensitivity to the torsion parameter space 
 is reduced even more. This happens for the following reason.
 The rate of $t\bar{t}$ signature from torsion production
 does not change
 with the increase of  $\eta$ while  $\eta_t$ is constant,
 since the increase of the production is compensated by the decrease of
 $Br(TS\to t\bar{t})$. In the same time the {\it width}
 of the torsion got increased. 
This increase of the torsion width causes the reduction of the signal 
significance and the respective reduction of  the LHC  sensitivity to the  
torsion parameters space. 
 One can see this clearly in  Fig.~\ref{fig:pp_tt}d
 demonstrating the absence of the LHC sensitivity
 to torsion parameter space
 for $\eta\gtrsim 0.3$ and $\eta_t=0.1$.
 Contrary, for $\eta_t>\eta$ case (Fig.~\ref{fig:pp_tt}b)
 the sensitivity of  the LHC to the torsion parameter space
 is increased.

\section{Conclusions}
  
We have found that LHC collider can establish unique constraints 
on the interactions between fermions and torsion field considerably
exceeding present experimental bounds.

Due to the renormalization group running, the universal interaction 
between all fermions and torsion at the Planck scale may be, in 
principle, characterized by the different values of the fermion-torsion 
couplings at the TeV scale. This effect should take place for all kinds 
of fermions. However, the numerical effect of the running is negligible 
for all known fermions except the top quark. Therefore, the parameter 
space of the torsion-fermion interactions is effectively reduced to the  
3-dimensional one, namely $\,(M_{TS},\eta,\eta_t)$.
Our phenomenological analysis has shown that this parameter space can 
be severely constrained with Drell-Yan and $\,t\bar{t}\,$ processes at 
the LHC. Moreover, for the first time we have demonstrated how possible 
non-universal nature of torsion-fermion interactions
can be tested via the combined analysis of Drell-Yan
and  $t\bar{t}$ processes.
The results are summarized in Fig.~\ref{fig:pp_ee}
and Fig.~\ref{fig:pp_tt} demonstrating
that LHC can improve present bounds
on the torsion parameters by about one order of magnitude.
For example, for $\eta=0.1$,  $M_{TS}\lesssim 4.5$~TeV
can be excluded at 95\%~CL.
However the test of the non-universal nature of 
$\eta\simeq\eta_t\simeq 0.1$ requires $M_{TS}\simeq 2$~TeV
which provides high enough statistics
for both
 $pp\to TS \to \ell^+\ell^-$ and
 $pp\to TS \to \bar{t}\bar{t}$ processes.

We have also shown that, since torsion width can be very large,
the four-fermion contact interaction approach used previously in 
several publications, could be not a good approximation for the 
investigation of the torsion phenomenology. 
Finally, we would like to mention that there is a room 
for more sophisticated analysis and combination of
electron and muon channels to improve further the LHC reach
for the torsion parameter space. Also, more sophisticated 
analysis with full detector simulation of the $pp\to t\bar{t}$ 
process and   $t\bar{t}$ pair reconstruction are necessary to 
understand more precisely the LHC sensitivity to non-universal 
nature of $\eta$ and $\eta_t$ couplings. We consider this  
paper as the first step in bringing ideas on the torsion 
phenomenology at the LHC to HEP community to stimulate further 
studies at the LHC.


\section*{Acknowledgments.} 

The work of M.A.B. do Vale and I.Sh. has been partially 
supported by the PRONEX project, research grants from 
FAPEMIG (MG, Brazil) and CNPq (Brazil). Besides, I.Sh. 
was supported by the Senior Associated Membership Grant 
from ICTP (Italy). The work of A.B. was supported by the 
US National Science Foundation under award PHY-0555545.
I.Sh. is also grateful to the Theory Division of CERN and 
Department of Theoretical Physics at the University of 
Zaragoza for kind hospitality. 


\end{document}